\documentclass[aps,prd,twocolumn,showpacs]{revtex4}

\usepackage{amsmath}
\usepackage{amsfonts}
\usepackage{graphicx}

\begin{document}

  \title{Can there be a heavy sbottom hidden in three-jet data at LEP?}
  \author{Rahul Malhotra }
  \email[Email: ]{rahul@math.utexas.edu}
  \affiliation{Center for Particle Physics, University of Texas, 
Austin, Texas, 78712, U.S.A.}
  \date{16 November, 2003}
  
  \begin{abstract}
A low-energy supersymmetry scenario with a light gluino of mass 
$12 - 16$ GeV and light sbottom ($\tilde{b}_1$) of mass 
$2 - 6$ GeV has been used to explain the apparent overproduction 
of $b$ quarks at the Tevatron. In this scenario the other mass eigenstate 
of the sbottom, i.e. $\tilde{b}_2$, is favored to be lighter 
than $180$ GeV due to constraints from electroweak 
precision data. We survey its decay modes in this scenario and 
show that decay into a $b$ quark and gluino should be dominant. Associated 
sbottom production at LEP via 
$e^+ e^- \rightarrow Z^* \rightarrow \tilde{b}_1 \tilde{b}^*_2 + 
\tilde{b}^*_1 \tilde{b}_2$ is studied and we show that it is naturally 
a three-jet process with a small cross-section, increasingly 
obscured by a large Standard Model background for heavier $\tilde{b}_2$. 
However we find that direct observation of a $\tilde{b}_2$ at the 
$5\sigma$ level is possible if it is lighter than $110 - 129$ GeV, depending 
on the sbottom mixing angle $|\cos\theta_b| = 0.30 - 0.45$. 
We also show that $\tilde{b}_2$-pair production can be mistaken for 
production of neutral MSSM Higgs bosons in the channel 
$e^+ e^- \rightarrow h^0 A^0 \rightarrow b\bar{b}b\bar{b}$. 
Using searches for the latter we place a lower mass limit of $90$ GeV on 
$\tilde{b}_2$.
  \end{abstract}
  \maketitle

  \section{Introduction}
The Standard Model (SM) has been very successful in explaining a range 
of observations at hadron colliders and the CERN $e^+ e^-$ collider, LEP. But 
it is still widely believed to be an effective theory valid at the 
electroweak scale, with new physics lying beyond it. The Minimal 
Supersymmetric Standard Model (MSSM)\cite{haber} is widely considered to 
be the most promising candidate for physics beyond the SM.

The MSSM contains supersymmetric (SUSY) partners of quarks, gluons and 
other SM particles which have not been observed, leading to speculation 
that they might be too heavy to have observable production rates at present 
collider energies. However it has been suggested in \cite{lightsb} that 
a light sbottom ($\tilde{b}_1$) with mass $\cal{O}$($5$ GeV) is not ruled out 
by electroweak precision data if its coupling to the $Z$ boson is tuned to be 
small in the MSSM. Recently Berger {\it et al} \cite{berger1} have also 
proposed a light sbottom and light gluino (LSLG) model to explain 
the long-standing puzzle of overproduction of $b$ quarks at the 
Tevatron \cite{excessb}. 
In this model gluinos of mass $12 - 16$ GeV are produced in pairs in $p
\bar{p}$ collisions and decay quickly into a $b$ quark and light sbottom 
($2 - 6$ GeV) each. The sbottom evades direct detection by quickly undergoing 
$R$-parity violating decays into soft dijets of light quarks 
around the cone of the accompanying $b$ jet. The extra 
$b$ quarks so produced result in a remarkably 
good fit to the measured transverse 
momentum distribution $\sigma_b (p_T > p^{min}_T)$ at NLO level, including 
data enhancement in the $p^{min}_T \sim m_{\tilde{g}}$ region.

Some independent explanations within the SM have also 
been proposed to resolve the discrepancy.  These include 
unknown NNLO QCD effects, updated $b$-quark 
fragmentation functions \cite{cacciari} and effects from changing the 
renormalization scale \cite{chyla}. But, without 
an unambiguous reduction in theoretical and experimental errors, 
the LSLG scenario cannot be ruled out. It is also interesting in its 
own right even if not solely responsible for the Tevatron discrepancy. 
For example, a light $\tilde{b}_1$ is more natural if the gluino 
is also light \cite{dreiner}. Experimental bounds on light gluinos 
do not apply here as either the mass range or the decay channel is 
different: only gluinos lighter 
than $6.3$ GeV \cite{janot} are absolutely ruled out. 
Very recently ALEPH \cite{stablesb} has ruled out 
stable sbottoms with lifetime $\gtrsim 1$ ns and mass $< 92$ GeV. 
However, using formulae in \cite{berger2} we calculate that even 
minimal $R$-parity violating 
couplings, as small as $10^{-6}$ times experimental limits, 
would leave $\tilde{b}_1$ with a lifetime shorter than $1$ ns. 
Light gluino and sbottom contributions to the running 
strong coupling constant $\alpha_S$($Q$) have also been calculated and 
found to be small \cite{berger1,alpha_s}. New phenomenon such as 
SUSY $Z$-decays \cite{rahul,cheung1,luo} 
and gluon splitting into gluinos \cite{cheung2} are predicted 
in this scenario, but the 
rates are either too small or require more careful study of LEP data.

The sbottoms and light gluinos also affect electroweak precision 
observables through virtual loops. In this case, serious constraints 
arise on the heavier eigenstate of the sbottom, 
i.e. $\tilde{b}_2$. According to \cite{cao}, corrections to $R_b$ are 
increasingly negative as $\tilde{b}_2$ becomes heavier and it has to 
be lighter than $125$ ($195$) GeV at the $2\sigma$ 
($3\sigma$) level. An extension of this analysis to the 
entire range of 
electroweak precision data \cite{cho} yields that 
$\tilde{b}_2$ must be lighter than $180$ GeV at $5\sigma$ level. 
However, it has been suggested that the SUSY decay 
$Z \rightarrow \tilde{b}_1 \bar{b} \tilde{g} + h.c.$ can contribute 
positively to $R_b$ \cite{cheung1}, reducing some of the negative loop 
effects, and possibly allowing higher $\tilde{b}_2$ masses 
\cite{rahul, luo2}. Independently, if large $CP$-violating phases 
are present in the model a $\tilde{b}_2$ with mass $\gtrsim 200$ GeV 
is possible \cite{baek}. Still, it is fair to say that in the face 
of electroweak constraints the LSLG model at least favors a 
$\tilde{b}_2$ lighter than $200$ GeV or so. 

In this article we study production and decay of such a heavy sbottom at 
LEPII. Available channels are (i) 
pair production: 
$e^+ e^- \rightarrow \tilde{b}_2 \tilde{b}^*_2$ and (ii) 
associated production: $e^+ e^- \rightarrow \tilde{b}^*_1 \tilde{b}_2 + 
\tilde{b}_1 \tilde{b}^*_2$. With LEPII center-of-mass energies 
ranging upto $\sqrt{s} = 209$ GeV, the second channel should have produced 
heavy sbottoms with masses as high as $\sim 200$ GeV. Since 
they have not been observed, it has been commented 
that the LSLG scenario is disfavored \cite{cao,cho}.

However, searches for unstable sbottoms at LEPII have not been done for the 
decay $\tilde{b}_2 \rightarrow b\tilde{g}$, which should dominate in 
this scenario as squarks, quarks and gluinos have strong 
trilinear couplings in the MSSM. In that case, 
the fast-moving gluino emitted by 
$\tilde{b}_2$ would decay quickly into a $b$ quark and $\tilde{b}_1$ that 
are nearly collinear, with 
$\tilde{b}_1$ subsequently undergoing $R$-parity violating decays 
into light quarks around the cone 
of the accompanying $b$-jet. Unless the jet resolution is set very high, 
the gluino should look like a fused $b$ flavored jet. 
Overall $\tilde{b}_2$ should appear as a heavy particle decaying 
into $b$ flavored dijets. On the other hand, the highly boosted prompt 
$\tilde{b}_1$ produced in the 
associated process would decay into nearly collinear light quarks 
and appear as a single hadronic jet. 
Pair and associated production are therefore naturally 
described as 4-jet and 3-jet 
processes respectively at leading order. Pair production in particular should 
be similar to neutral MSSM Higgs production in the channel 
$e^+ e^- \rightarrow h^0 A^0 \rightarrow b\bar{b}b\bar{b}$ if 
$h^0$ and $A^0$ have approximately equal masses.

The article is organised as follows: $\tilde{b}_2$ decays are studied 
in Section II and $\tilde{b}_2 \rightarrow b\tilde{g}$ is found to be 
dominant, cross-sections and event topology are studied in Section III and the 
corresponding SM 3-jet background for associated production is studied 
in Section IV. In Section V, LEP searches for neutral Higgs bosons 
are used to derive a lower bound on $\tilde{b}_2$ mass. 
Conclusions are drawn in Section VI.

  \section{Heavy Sbottom Decay}

Sbottom decays in MSSM scenarios with large mass splitting between 
$\tilde{b}_2$ and $\tilde{b}_1$ have been investigated before; 
see \cite{porod} for example. However 
the scenario where the gluino is also light has not received much attention. 

The direct decay products can be purely 
fermionic (1) or bosonic (2): 
\begin{eqnarray}
\tilde{b}_2 & \rightarrow & b\tilde{g}, b\chi^0_k, t\chi^- \label{eqn1}\\
\tilde{b}_2 & \rightarrow & \tilde{b}_1 Z, \tilde{t}W^-, 
\tilde{b}_1 h^0, \tilde{b}_1 A^0, \tilde{b}_1 H^0, \tilde{t}H^- \label{eqn2}
\end{eqnarray} 
where $\chi^0_k$ $(k = 1,..,4)$ and $\chi^{\pm}$ are 
neutralinos and charginos respectively, $t$ is the top quark, 
$\tilde{t}$ are stops, 
$h^0$ and $H^0$ are neutral $CP$-even 
Higgs bosons, $A^0$ is the $CP$-odd Higgs and $H^{\pm}$ are charged Higgs 
bosons. 

The individual widths depend on masses of above particles, but available 
experimental constraints \cite{PDG} are model-dependent and might not all 
be applicable in the LSLG scenario. However precision $Z$-width measurements 
can be used to apply some basic constraints on masses and 
the sbottom mixing angle.

\subsection{Couplings and mass constraints}

In the MSSM, $Z$-boson couplings to sbottom pairs are given by,
\begin{eqnarray}
Z\tilde{b}_1 \tilde{b}_1 & \propto & \frac{1}{2}\cos^2 \theta_b - 
\frac{1}{3}\sin^2 \theta_W \label{eqn3}\\
Z\tilde{b}_1 \tilde{b}_2 & \propto & -\frac{1}{2}\sin\theta_b\cos\theta_b 
\label{eqn4}\\
Z\tilde{b}_2 \tilde{b}_2 & \propto & \frac{1}{2}\sin^2 \theta_b - 
\frac{1}{3}\sin^2 \theta_W \label{eqn5}
\end{eqnarray}
where $\theta_b$ is the mixing angle between left and right-handed states:
\begin{equation}
\begin{pmatrix}
  \tilde{b}_1 \\
  \tilde{b}_2
\end{pmatrix}
 = 
\begin{pmatrix}
  \cos \theta_b &
  \sin \theta_b \\
  - \sin \theta_b &
  \cos \theta_b
\end{pmatrix}
\begin{pmatrix}
  \tilde{b}_L \\
  \tilde{b}_R
\end{pmatrix}\label{eqn6}
\end{equation}
The light sbottom should have a vanishingly small coupling in 
Eqn. (\ref{eqn3}) as the $Z \rightarrow \tilde{b}_1 \tilde{b}^*_1$ decay 
does not occur to high accuracy. 
This is achieved with the choice
\begin{eqnarray}
\cos \theta_b & \approx & \pm \sqrt{\frac{2}{3}}\sin\theta_W = \pm 0.39.
\label{eqn7}
\end{eqnarray}
The narrow range $|c_b| = 0.30 - 0.45$ ($c_b \equiv \cos\theta_b$) 
is allowed \cite{lightsb} which we use at times to obtain upper and 
lower bounds.

Given that $m_{\tilde{b}_1} = 2 - 6$ GeV, the decay 
$Z \rightarrow \tilde{b}_1 \tilde{b}^*_2 + h.c.$ might also take place 
if $\tilde{b}_2$ is lighter than $\sim 89$ GeV. However this decay is 
suppressed both kinematically and by the factor $\sin^2 2\theta_b$. 
Even for the higher value 
$|c_b| = 0.45$ we calculate $\Gamma 
(Z \rightarrow \tilde{b}_1\tilde{b}^*_2 + h.c.) \leq 10$ MeV for 
$m_{\tilde{b}_2} \geq 55$ GeV and $m_{\tilde{b}_1} \geq 2$ GeV. With the 
full $Z$-width having a $1\sigma$ error of $2.3$ MeV and a $0.6\sigma$ pull 
from the theoretical SM calculation \cite{PDG}, a lower limit of 
$55$ GeV 
on $\tilde{b}_2$-mass can be set at $\sim 4\sigma$ level without a detailed 
analysis.

Similarly, decays into pairs of neutralinos, charginos and stops 
might contribute unacceptably to the $Z$ width and it seems safe enough to 
apply a lower mass limit of $M_Z/2$ to them for calculation purposes. 
With the observed top quark mass of $\sim 175$ GeV, this rules out the 
chargino channel $\tilde{b}_2 \rightarrow t\chi^-$ as 
$\tilde{b}_2$ masses $\lesssim 200$ GeV are being considered. 

\subsection{Calculations}

The decay width for $\tilde{b}_2 \rightarrow b\tilde{g}$ is 
easily calculated at tree-level using Feynman rules for 
the MSSM given in \cite{rosiek}:
\begin{eqnarray}
\Gamma (\tilde{b}_2 \rightarrow b\tilde{g}) & = &
\frac{g^2_s m_{\tilde{b}_2} \kappa A}{6\pi}, \label{eqn8}\\
A & = & 
1 - x^2_b - x^2_{\tilde{g}} - 2 x_b x_{\tilde{g}} \sin 2\theta_b \nonumber
\end{eqnarray}
where $x_i = \frac{m_i}{m_{\tilde{b}_2}}$, 
$\kappa^2 = \sum_i x^4_i - \sum_{i \neq j}x^2_i x^2_j$ (summing 
over all particles involved in the decay) is the usual kinematic factor and 
$g_s$ is the strong coupling evaluated at 
$Q = m_{\tilde{b}_2}$. The canonical strong coupling value 
$\alpha_S (M_Z) = 0.118$ is used here. Other parameters used in this section 
are $m_b = 4.5$ GeV, 
$m_{\tilde{b}_1} = 4$ GeV, $m_{\tilde{g}} = 14$ GeV and $c_b = +0.39$. 

\begin{figure}[t!]
\includegraphics{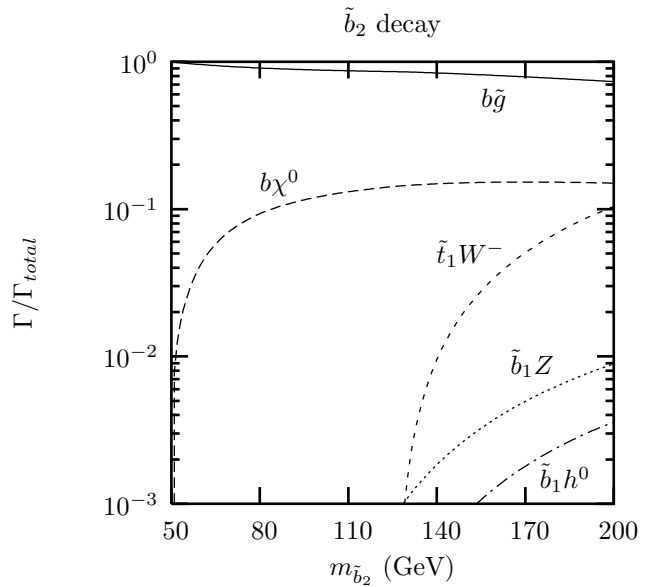}
\caption{\small \label{fig1}
Branching ratios for $\tilde{b}_2$ with $\tan\beta = 30$. Masses 
are set as $m_{\chi^0_k},m_{\tilde{t}_1} = M_Z/2$ $\forall$ $\chi^0_k$ and 
$m_{h^0} = 114.4$ GeV. The Higgs width is calculated in the decoupling limit.}
\end{figure}

The remaining widths in Eqns. (\ref{eqn1},\ref{eqn2}) are 
calculated using tree-level formulae 
given in \cite{porod}. Fig. \ref{fig1} shows the branching ratios versus 
$\tilde{b}_2$ mass. 
The $b\tilde{g}$ width is large, varying between $3.9 - 13.8$ GeV for 
$m_{\tilde{b}_2} = 55 - 200$ GeV. It has 
the maximum amount of available phase space and 
proceeds via the strong coupling, 
while the other widths are $\propto g^2_w$ where $g_w = e/\sin\theta_W$ 
is the usual weak coupling. 

The width shown for $\tilde{b}_2 \rightarrow 
b\chi^0$ is the summed width over all 4 neutralinos ($\chi^0_k$). 
This value scales approximately as 
$m^2_b \tan^2\beta$ for large $\tan\beta$. Here $\tan \beta = v_2/v_1$ 
where $v_i$ are the vacuum expectation values of the two Higgs doublets. 
Our calculation is most likely an overestimate as mixing angles are ignored 
and all neutralinos are prescribed 
the same mass. This channel has 
been extensively searched for at LEP \cite{sbsearch}, but seems to be 
at most $10 - 15\%$ of the full width in the LSLG scenario. 

Bosonic decays with $W$, $Z$ in the final state are also found to be small. 
We show $\Gamma(\tilde{b}_2 \rightarrow \tilde{t}_1 W^-)$ correct upto an 
unknown factor $\sin^2 \theta_t \leq 1$ where $\sin\theta_t$ is the 
stop mixing angle. For $\tilde{t}_2$ the factor would be $\cos^2 \theta_t$. 
Because of the unnaturally low value of $\tilde{t}_1$ mass chosen here, this 
width rises significantly as $m_{\tilde{b}_2}$ approaches $200$ GeV.

Decays into Higgs bosons are more complex as besides Higgs masses, 
the widths depend on unknown soft SUSY-breaking mass terms $A_b$ and $\mu$. 
The only available mass constraint is $m_{h^0} \lesssim 130$ GeV 
at two-loop level in the MSSM. However the excellent agreement between 
electroweak precision measurements and theoretical 
predictions with a single SM Higgs 
boson has led to a preference for the ``decoupling limit'' of 
the MSSM Higgs sector. In this limit, Yukawa 
couplings of $h^0$ to quarks and leptons are nearly identical to 
those of the Standard Model Higgs. At the same time $A^0,H^0,H^{\pm}$ 
have almost degenerate masses $>> M_Z$. Therefore, 
with $\tilde{b}_2$ 
lighter than $200$ GeV, only $\tilde{b}_2 \rightarrow \tilde{b}_1 h^0$ 
is likely to be significant while other decays would be kinematically 
impossible or heavily suppressed. The width is then given by
\begin{eqnarray}
\Gamma (\tilde{b}_2 \rightarrow \tilde{b}_1 h^0) & = & 
\frac{g^2_w \kappa B^2}{64\pi m_{\tilde{b}_2}}, \label{eqn9}\\ 
B & = &  - \frac{m_b \cos 2\theta_b}{m_W}(A_b - \mu \tan \beta) \nonumber\\
& & +\frac{m_Z \sin 2\theta_b}{\cos \theta_W}(-\frac{1}{2} + 
\frac{2}{3}\sin^2 \theta_W)\cos 2\beta \nonumber
\end{eqnarray}
We choose $m_{h^0} = 114.4$ GeV in our calculation as LEP data has ruled out 
SM Higgs bosons lighter than this value \cite{smhiggs}.

In the decoupling limit, 
arbitrary variation over $A_b$, $\mu$ in calculating $B$ is not required 
as the factor $A_b - \mu \tan\beta$ can be expressed in terms of sbottom 
masses and $\theta_b$:
\begin{eqnarray}
\sin 2\theta_b & = & \frac{2m_b (A_b - \mu \tan \beta)}{m^2_{\tilde{b}_1} 
- m^2_{\tilde{b}_2}}\label{eqn10}
\end{eqnarray}
with $\theta_b$ given by Eqn. (\ref{eqn7}). 
This is a common relation that arises when the sbottom mass matrix 
(see \cite{primer} for example) is 
diagonalized with the mixing matrix in Eqn. (\ref{eqn6}).

Though theoretically and experimentally attractive, 
if the decoupling limit does not 
hold then other 
Higgs particles might also be light. The most general lower mass limits from 
LEP on neutral MSSM Higgs bosons are about $90$ GeV \cite{higgssearch}. 
Then, the 
$\tilde{b}_2 \rightarrow 
\tilde{b}_1 A^0$ width (say) can become larger than $10\%$ of $\tilde{b}_2 
\rightarrow b\tilde{g}$ due to the 
coupling
\begin{eqnarray}
A^0\tilde{b}_1\tilde{b}_2 & \propto & -\frac{g_w m_b\cos 2\theta_b}
{2m_W}(\mu + A_b\tan\beta)\label{eqn11}
\end{eqnarray}
This happens if $A_b\tan\beta$ is larger than $\sim 10$ TeV. 
Though the possibility is there, we consider it less likely and do not 
pursue it further. In any event such a decay would be more important for 
higher $\tilde{b}_2$ masses, and we show in Section III that 
$\tilde{b}_2$ production at LEPII falls rapidly as its mass nears 
$200$ GeV.

We therefore conclude that the strong decay 
$\tilde{b}_2 \rightarrow b\tilde{g}$ is dominant and other decays 
are unlikely to be of more than marginal importance at LEPII. 
 
\section{Production at Lep}
Cross-sections for $\tilde{b}_2$ production are defined as 
follows: 
$\sigma_{22} = \sigma (e^+ e^- \rightarrow \tilde{b}_2 \tilde{b}^*_2)$ 
and 
$\sigma_{12} = \sigma (e^+ e^- \rightarrow \tilde{b}_1 \tilde{b}^*_2 + h.c.)$. 
For completeness production of $\tilde{b}_1 \tilde{b}^*_1$ pairs 
is referred to 
as $\sigma_{11}$. The $\sigma_{ij}$ are readily 
calculated at tree-level,
\begin{eqnarray}
\sigma_{ij} & = & \frac{g^4_w \sin^4 \theta_W \beta^3_{ij}}{16\pi s}f_{ij}, 
\label{eqn12}\\
f_{ij} & = & (\frac{1}{9}-\frac{2c_V \lambda_{ij}}{3\beta^2_Z\sin^2 2\theta_W})
\delta_{ij}+\frac{(c^2_V+c^2_A)\lambda^2_{ij}}{\beta^4_Z\sin^4 2\theta_W}
\label{eqn13}
\end{eqnarray}
where $\lambda_{11} \approx 0$, 
$\lambda_{12} = \frac{1}{\sqrt{2}}\sin 2\theta_b$, 
$\lambda_{22} 
= \sin^2 \theta_b - \frac{2}{3}\sin^2 \theta_w$, $\beta^2_{ij} = 
(1 - \frac{(m_{\tilde{b}_i}+m_{\tilde{b}_j})^2}{s})(1 - 
\frac{(m_{\tilde{b}_i}-m_{\tilde{b}_j})^2}{s})$, 
$\beta^2_Z = 1 - \frac{M^2_Z}{s}$ 
and $c_{V,A}$ are electron vector and axial couplings that equal 
$-\frac{1}{2}+2\sin^2 \theta_W$ and $\frac{1}{2}$ respectively. The 
$\lambda$-factors are proportional to sbottom-$Z$ couplings 
in Eqns. (3-5). We use the same parameters here as used earlier 
for width calculations.

Both virtual photon ($\gamma^*$) and virtual $Z$ ($Z^*$) channels are 
available for $\sigma_{22}$ while only $Z^*$ 
is available for $\sigma_{12}$. The latter falls by a factor of 
$2$ in going from $|c_b| = 0.45$ to $0.30$. Pair production rises 
in the same range by a smaller factor of 1.3 at $\sqrt{s} = 207$ GeV. 
Variation of $\tilde{b}_1$ mass between $2 - 6$ GeV has 
negligible effect on $\sigma_{12}$.

\begin{figure}[t!]
\includegraphics{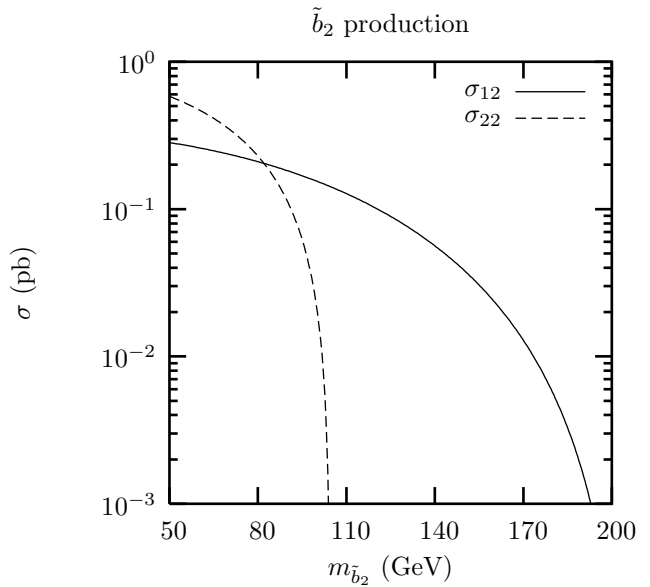}
\caption{\small \label{fig2}
The $\tilde{b}_2$ production cross-section for $\sqrt{s} = 207$ GeV, 
$|c_b| = 0.39$ as a function of mass.} 
\end{figure}

Fig. \ref{fig2} shows $\sigma_{ij}$ versus $\tilde{b}_2$-mass at 
$\sqrt{s} = 207$ GeV. Both 
cross-sections are suppressed due to the $\beta^3$ kinematic factor 
for scalar particle 
production. However, asymmetry between sbottom masses causes 
additional kinematic suppression of $\sigma_{12}$ as $\beta_{12} 
\approx \beta^2_{22}$ for the same total rest mass of final products, 
$m_{\tilde{b}_i} + m_{\tilde{b}_j}$. The missing photon 
channel and smaller $\lambda$-factor, 
$\lambda^2_{22}/\lambda^2_{12} \approx 1.8$, 
reduces the cross-section further. Therefore 
associated production is generally small and falls rapidly 
as $\tilde{b}_2$ gets heavier. 

The LEPII operation covered a range 
of center-of-mass energies from $130 - 209$ GeV with maximum data 
collected at $\sqrt{s} = 189$ GeV and $205 - 207$ GeV. 
Fig. \ref{fig3} shows the expected number of raw events. We use an approximate 
luminosity distribution provided in \cite{lepewwg} counting 
the combined integrated luminosity recorded by all four LEP experiments. 
The number of events for associated production falls below $\sim 100$ for 
$m_{\tilde{b}_2} > 147$ GeV at $|c_b| = 0.39$. It is therefore 
possible that sufficient statistics might not be available to 
explore sbottom masses above this value.

\begin{figure}[b!]
\begin{tabular}{|c||c|c|}
\hline
{\it Number} & \multicolumn{2}{|c|}{{\it Maximum} $\tilde{b}_2$ {\it mass (GeV)
}} \\
\cline{2-3}
{\it of} & {\it Associated} & {\it Pair} \\
{\it Events at LEPII} & {\it Production} & {\it Production} \\
\cline{1-1} \cline{2-3}
1000 & 59 & 71 \\
100 & 147 & 94 \\
10 & 177 & 101 \\
1 & 192 & 103 \\
\hline
\end{tabular}
\caption{\small \label{fig3}
Expected number of raw LEPII events for the combined luminosity 
recorded in the entire run. We show the $\tilde{b}_2$ 
masses beyond which event counts fall below rough benchmark levels.}
\end{figure}

\begin{figure}[t!]
\includegraphics{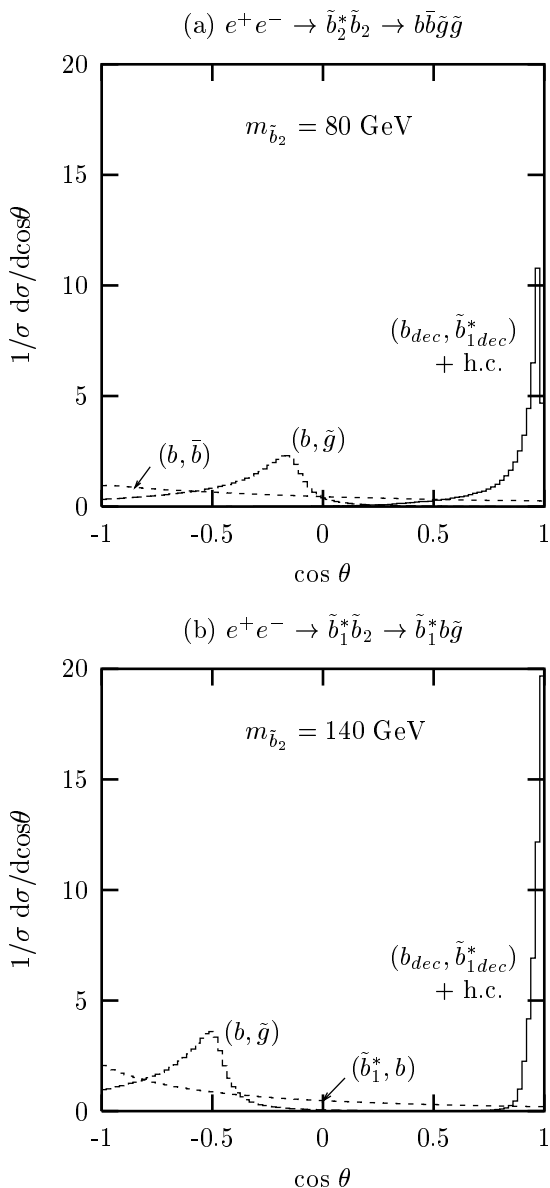}
\caption{\small \label{fig4}
Opening angles between particle pairs in (a) pair production and 
(b) associated production at $\sqrt{s} = 207$ GeV. Particles  marked with 
``$dec$'' are gluino decay products. In (a) the $(b,\tilde{g})$ distribution 
shown is for $b$ quarks and gluinos arising from the same $\tilde{b}_2$. 
$(b,\tilde{g})$ arising from different $\tilde{b}_2$ and 
$(\tilde{g},\tilde{g})$ have an identical distribution to that 
shown for $(b,\bar{b})$. In (b), $(\tilde{b}^*_1,\tilde{g})$ is not 
shown as it is the same as $(\tilde{b}^*_1,b)$.} 
\end{figure}

We now discuss the event topology in order to identify 
important backgrounds. As shown in Section II the decay 
$\tilde{b}_2 \rightarrow b\tilde{g}$ is dominant 
which results in the states $\tilde{b}_1 \bar{b}\tilde{g} + h.c.$ 
and $b\bar{b}\tilde{g}\tilde{g}$ for associated and pair processes 
respectively. We decay the gluinos into $b\tilde{b}^*_1/\bar{b}
\tilde{b}_1$ pairs and show the opening angles between final 
products for some representative $\tilde{b}_2$ masses in Fig. \ref{fig4}. 
The $b$ quark and $\tilde{b}_1$ arising 
from gluino decay overwhelmingly prefer a small angular 
separation with a sharp peak at $\cos\theta \gtrsim 0.9$. 
The other particles tend to be well-separated. 

Through $R$-parity and baryon-number violating couplings 
$\lambda^{''}_{ij3}$, $\tilde{b}_1$ can decay into 
pairs of light quarks: $\tilde{b}^*_1 \rightarrow u + s; c + d; c + s$. 
A detailed discussion of such decays is 
given in \cite{berger2}. In that case, the $\tilde{b}_1$ arising from 
gluino decay would further decay hadronically in and around the cone of 
the accompanying $b$ jet. In practise it would be 
difficult to distinguish between the overlapping jets, unless a very fine 
jet resolution is used. The gluino should then appear for the most part 
as a single fused $b$-flavored jet with perhaps 
some extra activity around the cone. 

The prompt $\tilde{b}_1$ from associated production is highly boosted 
for most $\tilde{b}_2$ masses within range. This should result in a very 
small angular separation between its decay products. If it decays into 
pairs of light quarks, we calculate that at 
$\sqrt{s} = 207$ GeV, $m_{\tilde{b}_1} = 4$ GeV 
and $m_{\tilde{b}_2} \lesssim 170$ GeV, 
at least $90\%$ of these would have 
an opening angle $< 30^{o}$. At any rate a $\tilde{b}_2$ as heavy as 
$170$ GeV is unlikely to be observable because of low event counts and 
would be obscured by the large 3-jet SM background (Section IV). 
Therefore in the observable range 
$\tilde{b}_1$ should show up as a single hadronic jet.

At leading order then, associated production is best described 
as a 3-jet process, with 2 jets that can be tagged as $b$ quarks 
and a hadronic jet from $\tilde{b}_1$. The relevant background for 
this would be SM 3-jet events which we discuss in 
Section IV. 
On the other hand, pair production is naturally 
a 4-jet process where each jet can be tagged as a $b$ quark. This 
would have significant background from 
{\it any other} heavy particles produced in pairs and decaying into 
dijets of $b$ quarks. Searches for neutral Higgs bosons $h^0$ and $A^0$ 
that can satisfy this criteria have been done, and we discuss them 
in Section V.

\section{3-jet Background}

The SM gluon radiation process: $e^+ e^- \rightarrow 
q\bar{q}g$, $q = u,d,s,c,b$; constitutes the 
main 3-jet background for associated production. In particular, 
$e^+ e^- \rightarrow b\bar{b}g$ could be an irreducible background as gluon 
jets and jets from light sbottoms might not be distinguishable on a 
case-by-case basis. 

We compare this background with associated production 
using the JADE jet-clustering algorithm 
\cite{jade}:
\begin{eqnarray}
\min_{i \neq j}{(p_i + p_j)^2} & \geq & y_{cut}s
\end{eqnarray}
where $p_i$ are the momenta of the final state partons and $0 < 
y_{cut} < 1$ is the jet resolution parameter. 
As long as $y_{cut} > m^2_{\tilde{g}}/s \approx 3.4 - 5.9 \times 10^{-3}$ for 
$\sqrt{s} = 207$ GeV and $m_{\tilde{g}} = 12 - 16$ GeV, 
the hadronic decay products of $\tilde{g}$ 
and $\tilde{b}_1$ are clustered into single jets. 
We evaluate matrix elements 
at leading order and do not consider 
contributions to the SM 3-jet cross-section from final states with more 
than three partons. The renormalization scale is set at 
$Q = \sqrt{s}/2$ with $\alpha_S (M_Z) = 0.118$. 

\begin{figure}[b!]
\includegraphics{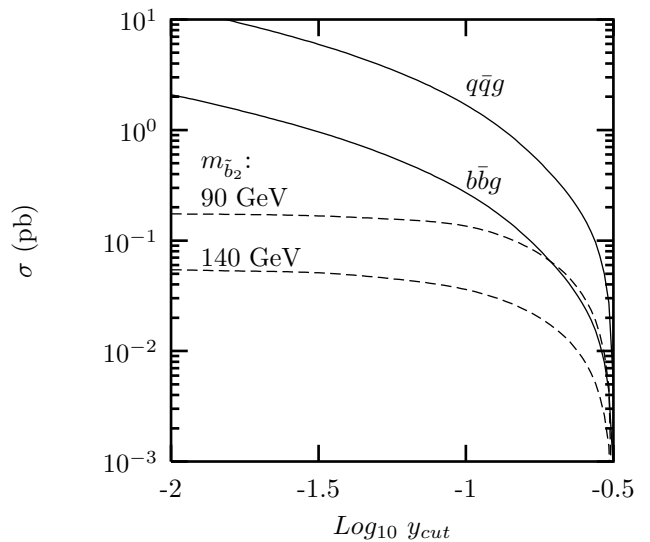}
\caption{\small \label{fig5}
Associated production (dashed lines) 
compared to SM 3-jet cross-sections versus $y_{cut}$ at 
$\sqrt{s} = 207$ GeV.}
\end{figure}

Fig. \ref{fig5} shows that $\sigma_{12}$ is a small fraction of the 
total SM 3-jet cross-section, though it increases in proportion 
as $y_{cut}$ increases and the jets are required to be well-separated. 
It is unlikely to be visible as a generic excess in 3-jet production 
given that measurements of hadronic cross-sections at LEPII have errors of 
at least $\pm 0.2$ pb \cite{lepewwg}. However, if at least one jet is 
$b$-tagged and $\sigma(e^+ e^- \rightarrow b\bar{b}g)$ is measured 
very accurately, then for $\tilde{b}_2$ lighter than $\sim 140$ GeV 
an excess might be observable at higher $y_{cut}$ values.

\begin{figure}[t!]
\includegraphics{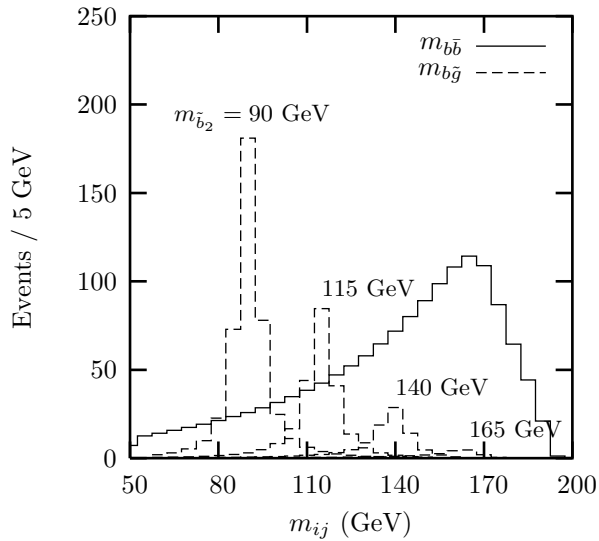}
\caption{\small \label{fig6}
The invariant mass of two $b$ tagged jets can be 
reconstructed to observe excesses. Dashed lines show 
associated production and the solid line $b\bar{b}g$ for 
$\log_{10} y_{cut} = -1.2$. Tagging efficiencies for $b$ quarks are 
not applied here. Events 
are shown for the total integrated luminosity 
recorded by the four LEP collaborations at $\sqrt{s} \geq 183$ GeV.}
\end{figure}

If two jets out of three are required to have $b$ tags 
then their total invariant mass can also be studied as in Fig. \ref{fig6}. 
The total invariant mass of the $b/\bar{b}$ quark and 
gluino (which appears as a 
$b$-like jet) gives rise to a clear resonance around 
$m_{\tilde{b}_2}$. This would allow direct observation of a $\tilde{b}_2$, 
and should be the preferred method of study.

The differential cross-section for $b\bar{b}g$ events increases with the 
invariant mass, $m_{b\bar{b}}$, while the resonance in 
$\sigma_{12}$ rapidly gets smaller as $\tilde{b}_2$ gets heavier. 
This is natural as gluon radiation from quark pairs is higher for 
softer gluons, which in turn implies a higher total invariant mass 
for the $b\bar{b}$ pair. To estimate the discovery region we calculate 
both signal ($S$) and background ($B$) events in the mass 
window $M_{bb} = m_{\tilde{b}_2} \pm \Delta M$ where $M_{bb}$ is the invariant 
mass of the $b$ tagged jets and $\Delta M = \Gamma_{\tilde{b}_2}$. The 
$b$ tagging efficiency $\epsilon_b$ is taken to be $65\%$, 
from $R_b$ studies at LEPII \cite{Rb}. 
Mistag probabilities are assumed to be small and 
not included in the analysis. We also use $\log_{10} y_{cut} = -1.2$ 
which is found to maximize the significance $S/\sqrt{B}$. The $N\sigma$ 
discovery region is defined as
\begin{eqnarray}
\frac{S}{\sqrt{S+B}} & \geq & N
\end{eqnarray}
Calculating events using the entire integrated luminosity recorded for 
$\sqrt{s} \geq 183$ GeV, we find that for $|c_b| = 0.39$, 
$\tilde{b}_2$ masses upto $123$ ($136$) GeV can be 
discovered at the $5\sigma$ ($3\sigma$) level. For $|c_b| = 0.30 - 0.45$, the 
upper limits for discovery are 
$m_{\tilde{b}_2} = 110 - 129$ GeV ($5\sigma$) and 
$m_{\tilde{b}_2} = 125 - 140$ GeV ($3\sigma$). Since $S$ and $B$ are 
$\propto \epsilon^2_b$, the significance is $\propto \epsilon_b$ and 
better $b$ tagging efficiencies can improve the upper limits. 
However we have not included effects of 
Gaussian smearing of pair invariant mass measurements, which might 
reduce the significance.
 
We note that the associated process also receives an irreducible 
SUSY background as the $\tilde{b}^*_1 b\tilde{g} + h.c.$ 
final state is possible even if the heavy sbottom is absent. This has 
been studied in the context of $Z$ decay \cite{cheung1}. 
However, its kinematics are very different from 
the same state produced by $\tilde{b}_2$ decay, and it should 
have little effect on the overall background. In Fig. 6 it 
would appear as an approximately uniform 
distribution of $\sim 5$ events $/ 5$ GeV, 
which is insignificant compared to the $b\bar{b}g$ background.

\section{Searches for $e^+ e^- \rightarrow h^0 A^0$}

At leading order $e^+ e^- \rightarrow h^0 A^0$ proceeds only through the 
virtual $Z$ channel. The relevant coupling is 
\begin{eqnarray}
Z h^0 A^0 & \propto & g_w \cos(\beta-\alpha)
\end{eqnarray} 
where 
$\alpha$ is the mixing angle between neutral $CP$-even 
Higgs bosons. This is comparable to the heavy sbottom coupling 
$Z\tilde{b}_2\tilde{b}_2 \propto g_w(\sin^2 \theta_b - 
\frac{2}{3}\sin^2\theta_W)$ in Eqn. \ref{eqn5}. However production of 
$\tilde{b}_2\tilde{b}^*_2$ pairs is somewhat higher as it also 
takes place through the $\gamma^*$ channel and receives an extra factor 
of $3$ from summing over final-state colors.

Being scalars, both pairs of particles are produced with the same 
angular distribution. Searches for $h^0 A^0$ production \cite{higgssearch} 
have been done 
along the diagonal $m_{h^0} = m_{A^0}$, which makes them 
kinematically identical to $\tilde{b}_2$ pair production. The final 
states searched for are 
$b\bar{b}b\bar{b}$, $b\bar{b}\tau^+\tau^-$ or $\tau^+\tau^-\tau^+\tau^-$ as 
$h^0/A^0$ decay mainly into $b$ or $\tau$ pairs in the parameter 
space where they are approximately equimassive. 
Therefore, the 4$b$ channel can be used to place limits on 
$\tilde{b}_2$ pair production as the latter leads to 4 
$b$ flavored jets in the final state.

\begin{figure}[t!]
\includegraphics{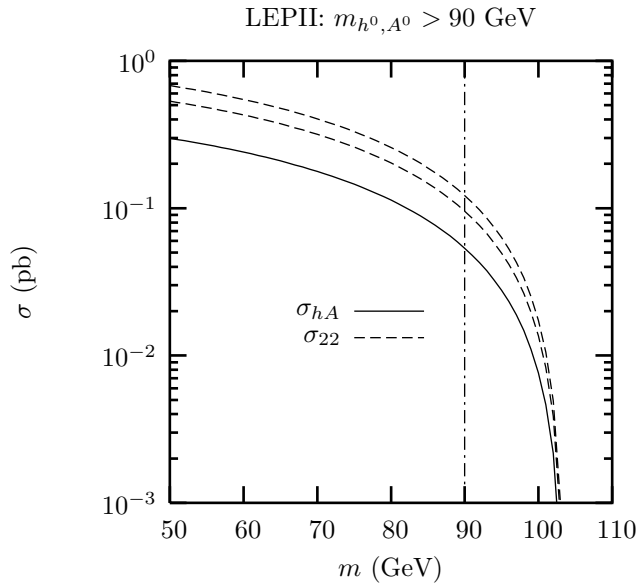}
\caption{\small \label{fig7}
Comparison between $\sigma_{hA}$ and $\sigma_{22}$ at 
$\sqrt{s} = 207$ GeV, versus $m = m_{h^0} = m_{A^0} = m_{\tilde{b}_2}$. 
Upper and lower limiting curves for $\sigma_{22}$ are 
obtained for $|c_b| = 0.30$, $0.45$ respectively.} 
\end{figure}

Cross-sections for the two processes are compared in Fig. \ref{fig7}. 
The $h^0 A^0$ cross-section is called $\sigma_{hA}$. 
We simply maximize this by setting $\cos (\beta-\alpha) = 1$ 
and Br$(h^0/A^0 \rightarrow b\bar{b}) = 1$. The parameters used in 
the experimental study were similar or lesser. We find that 
$\sigma_{22}$ is $1.8 -2.3$ times higher than Higgs 
production for $|c_b| = 0.45 - 0.3$. If the more typical 
branching ratios 
$Br(h^0 \rightarrow b\bar{b}) = 0.94$ and 
$Br(A^0 \rightarrow b\bar{b}) = 0.92$ are used then $\sigma_{22}$ is 
effectively 2.1 to 2.6 times higher. However that could be offset 
if $\tilde{b}_2$ has a branching ratio into $b\tilde{g}$ 
near its lower limit of around $0.9$ in this mass range 
(see Fig. \ref{fig1}).

Experimental searches for $h^0 A^0$ have used approximately $870$ pb$^{-1}$ 
of combined integrated luminosity, with center-of-mass energies between 
$200$ and $209$ GeV. Only OPAL has seen a significant excess 
in the 4$b$-jet channel, 
which is at the $2\sigma$ level at $(m_{h^0},m_{A^0}) \sim (93,93)$ GeV. 
This does not appear in other experiments, though it cannot be ruled out 
statistically. No excess in this channel 
seems to have been observed by any experiment below $\sim 90$ GeV which is 
approximately the quoted lower limit at $95\%$ confidence for Higgs masses. 
Since the pair cross-section is 
higher than that for $h^0 A^0$, this should simultaneously rule out 
heavy sbottoms lighter than $90$ GeV in the LSLG scenario.

There are some qualifications to this analysis. First, 
$\tilde{b}_2$ has a much larger width in absolute terms than $h^0$ or $A^0$, 
and that seems to have been a significant factor in the $h^0 A^0$ searches 
at LEP. However, since $\sigma_{22}$ is larger, 
it is likely that any excess would have been observed and the 
$90$ GeV lower limit is 
approximately correct. Secondly, if very low 
values of $y_{cut}$ (below $m^2_{\tilde{g}}/s$) were used in the LEP searches, 
then the above analysis might not hold.

\section{Conclusions}
We have shown that the heavy sbottom eigenstate decays dominantly into 
$b\tilde{g}$ pairs in the light sbottom and light gluino 
scenario. Pair and associated production of 
$\tilde{b}_2$ at LEPII have been studied and found to be 
naturally described as 4-jet and 3-jet processes respectively. 
Their cross-sections and raw event rates have been calculated and associated 
production is found to be small and obscured by the large SM 3-jet 
background for large values of $\tilde{b}_2$ mass. However, 
we find that $5\sigma$ discovery of a $\tilde{b}_2$ is possible  
using 3-jet data provided $m_{\tilde{b}_2} \leq 110 - 129$ GeV, 
for $|c_b| = 0.30 - 0.45$. The corresponding $3\sigma$ limits are 
$m_{\tilde{b}_2} \leq 125 - 140$ GeV. We recommend a search 
as far as possible. While invariant masses reconstructed from 
$b$-tagged jet pairs might be the most direct 
way to do this, single $b$-tagged events can also be useful 
if the cross-sections are measurable to a high accuracy. 

We also find that $\tilde{b}_2$ pair production is similar to 
production of neutral MSSM Higgs bosons decaying into $b\bar{b}$ pairs, which 
have been extensively searched for by the four LEP collaborations. 
Minor excesses, though 
inconclusive, seen in the $4b$ jet channel for 
masses $\sim 93$ GeV provide further motivation for a 
detailed study of 3-jet events. We show that $\tilde{b}_2$ should 
be heavier than about $90$ GeV as no excess has been reported below this value.

\section{Acknowledgements}

I would like to thank Prof. D. A. Dicus for useful discussions and help 
given throughout the course of this work. This work was supported in part 
by the United States Department of Energy under 
Contract No. DE-FG03-93ER40757.

{\it Note:} A paper by E.L. Berger, J. Lee and T.M.P. Tait 
(hep-ph/0306110) that also covers associated production in this scenario, 
using the jet cone algorithm, appeared independently on the internet a 
few days before this one.

\end{document}